\documentclass[journal]{IEEEtran}
\usepackage{cite}
\usepackage[numbers,sort&compress]{natbib}
\usepackage{amsmath,amssymb,amsfonts}
\usepackage{algorithmic}
\usepackage{graphicx}
\usepackage{textcomp}
\usepackage{xcolor}
\usepackage{tikz}
\usepackage[table]{xcolor}

\usepackage{amsmath}
\usepackage{amssymb}
\usepackage{soul}
\usepackage{url,subfigure,graphicx,wrapfig}
\usepackage{fixmath}
\usepackage{blindtext}
\usepackage{bm}
\usepackage[ruled,linesnumbered]{algorithm2e}
\usepackage{multirow}
\usepackage {xcolor}
\usepackage{pdfrender}
\usepackage{diagbox}
\usepackage{amsbsy}
\usepackage{booktabs}
\usepackage{threeparttable}
\usepackage{tabularx}
\usepackage{booktabs,tabularx,ragged2e,array}
\newcolumntype{Y}{>{\RaggedRight\arraybackslash}X}
\newcommand*\circled[1]{\tikz[baseline=(char.base)]{
            \node[shape=circle,draw,inner sep=0.3pt] (char) {#1};}}
\def\BibTeX{{\rm B\kern-.05em{\sc i\kern-.025em b}\kern-.08em
    T\kern-.1667em\lower.7ex\hbox{E}\kern-.125emX}}
\begin{document}

\title{Who's Wearing? Ear Canal Biometric Key Extraction for User Authentication on Wireless Earbuds}
\author{
\IEEEauthorblockN{Chenpei Huang,~\IEEEmembership{Student Member,~IEEE,}
Lingfeng Yao,~\IEEEmembership{Student Member,~IEEE,}
Hui Zhong,~\IEEEmembership{Student Member,~IEEE,}
Kyu In Lee,~\IEEEmembership{Member,~IEEE,}
Lan Zhang,~\IEEEmembership{Member,~IEEE,}
Xiaoyong Yuan,~\IEEEmembership{Senior Member,~IEEE,}\\
Tomoaki Ohtsuki,~\IEEEmembership{Senior Member,~IEEE,}
and Miao Pan, ~\IEEEmembership{Senior Member,~IEEE}}
\thanks{C. Huang, L. Yao, H. Zhong and M. Pan are with the Department of Electrical and Computer Engineering, University of Houston, Houston,
TX 77204, USA. (e-mail: chuang30@uh.edu; lyao12@uh.edu; hzhong5@uh.edu; mpan2@uh.edu).}
\thanks{K. Lee is with the Department of Information Science Technology, University of Houston, Houston,
TX 77204, USA. (e-mail: klee48@central.uh.edu).}
\thanks{L. Zhang and X. Yuan are with the Department of Electrical and Computer Engineering, Clemson University, Clemson, SC 29634, USA (e-mail: lan7@clemson.edu; xiaoyon@clemson.edu).}
\thanks{T. Ohtsuki is with Department of Information and Computer
Science, Keio University, Kanagawa 223-8522, Japan (e-mail: ohtsuki@keio.jp).}
}

\maketitle

\begin{abstract}
Ear canal scanning/sensing (ECS) has emerged as a novel biometric authentication method for mobile devices paired with wireless earbuds. Existing studies have demonstrated the uniqueness of ear canals by training and testing machine learning classifiers on ECS data. However, implementing practical ECS-based authentication requires preventing raw biometric data leakage and designing computationally efficient protocols suitable for resource-constrained earbuds.
To address these challenges, we propose an ear canal key extraction protocol, \textbf{EarID}. Without relying on classifiers, EarID extracts unique binary keys directly on the earbuds during authentication. These keys further allow the use of privacy-preserving fuzzy commitment scheme that verifies the wearer's key on mobile devices.
Our evaluation results demonstrate that EarID achieves a 98.7\% authentication accuracy, comparable to machine learning classifiers. The mobile enrollment time (160~ms) and earbuds processing time (226~ms) are negligible in terms of wearer's experience. Moreover, our approach is robust and attack-resistant, maintaining a false acceptance rate below 1\% across all adversarial scenarios. We believe the proposed EarID offers a practical and secure solution for next-generation wireless earbuds.
\end{abstract}


\section{Introduction}
Biometric authentication utilizes human physiological or behavioral characteristics to recognize users, offering a secure, reliable, and user-friendly authentication system. Due to their uniqueness across individuals, biometric methods only require a user to enroll their biometric data once for later authentication \cite{fingerpirnt, speakerVerification, faceRecognition}. Examples include fingerprint authentication on smartphones (TouchID \cite{AppleTouchID}), voice recognition for smart home control (Alexa Voice Control \cite{AlexaSmartHome}), and face-based online payment systems (Pay By Face \cite{PayByFace}). Recently, innovative biometric approaches leveraging advanced wearable technologies, such as wireless earbuds, smart glasses, and virtual reality headsets, have gained significant attention \cite{EarGait,ToothSonic,Voice-in-ear,EarPrint_OE,F2Key}. These wearables exhibit remarkable biometric sensing capabilities and introduce new requirements concerning usability for authentication \cite{metaverseAuthentication, meterverseCAN}.

\smallskip
\noindent \textbf{Biometrics using ear canal scanning.} \\
Offering an appealing ``hands-free" feature, wireless earbuds are particularly suitable for new biometric authentication designs \cite{ARsok}. Positioned inside the ear canal, ear-based authentication can: (1) passively capture signals such as tooth vibrations \cite{ToothSonic}, footsteps \cite{EarGait}, and bone-conducted audio \cite{Voice-in-ear}; and (2) actively sense otoacoustic emissions \cite{EarPrint_OE}, facial patterns \cite{SonicID} along with their dynamics \cite{F2Key}, and structural characteristics of the ear canal or middle ear \cite{2018inaudibleEar, 2019earecho, 2021eardynamic, 2021headfi, 2022bilateral, 2023earace}. Consequently, wireless earbuds can function effectively as biometric scanners, authenticating users for paired mobile devices, such as unlocking smartphones.
\begin{figure}
    \centering
    \includegraphics[width=\linewidth]{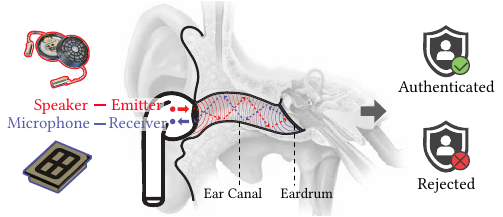}
    \caption{Illustration of ear canal scanning/sensing. The user inserts the earbuds to run ear canal scanning, acoustic sensing the unique ear canal structure, for biometric user authentication.}
    \label{fig:ear-scan}
\end{figure}
In this paper, we focus on a specific earable biometric: \textit{ear canal scanning/sensing (ECS)} for wireless earbuds. Typically, wireless earbuds are equipped with a speaker and inward-facing microphones. When placed inside the occluded ear canal, as illustrated in Fig.~\ref{fig:ear-scan}, they can emit acoustic energy into the ear canal and capture the resulting reflections, similar to the radar system. These reflection signals are shaped by the size and structure of the user's ear canal and middle ear, producing unique patterns that can be used to distinguish different individuals.

Recently, many user authentication schemes have been proposed using ECS biometrics. For instance, Mahto \textit{et al}~\cite{2018inaudibleEar} study the advantages of using human inaudible frequency ranges during the ear canal scanning. EarDynamic~\cite{2021eardynamic} operates at the inaudible frequencies, and exploits the ear canal deformation as extra discriminative features. HeadFi \cite{2021headfi} extends the in-ear sensing capacity to low-end devices without an in-ear microphone. While these studies focus on different aspects of ear-scanning technologies, they all employ machine learning classifiers trained on ECS data to make authentication decisions. 

\smallskip
\noindent \textbf{Wireless earbuds authentication challenges.}\\
Thanks to Bluetooth technology, wireless earbuds can be easily paired with mobile devices, enabling hands-free audio and voice interaction. With those salient features, wireless earbuds are gaining immense popularity among consumers, expecting a USD 563.2 billion market by 2030 \cite{yahooTWS2023}. Since wireless earbuds are becoming more powerful and versatile, one may ask: can we also develop wireless earbuds as biometric scanner for user authentication? Can we implement ECS on wireless earbuds? However, most existing studies rely on training machine learning classifiers, which presents several challenges for practical protocol design.

\smallskip
\noindent{\underline{\circled{1} Limited computation capacity.}}
Wireless earbuds have limited computational capacity due to their compact size, power efficiency requirements, and battery constraints. For instance, the microcontroller unit (MCU) in a typical wireless earbud operates at a frequency roughly ten times slower than that of a standard laptop processor\footnote{Cortex-M4 @ 160 MHz for earbuds \cite{nxpEarbuds} vs. 1.6 GHz laptop CPU. Other architectural factors also influence processing speed.}. However, existing ECS classifiers typically require training after user enrollment, which involves computationally intensive iterative optimization. Although some high-end earbuds support on-device machine learning, such capabilities are mainly reserved for tasks like active noise cancellation and voice command detection using lightweight inference models, not for biometric authentication. Conversely, if a binary key effectively captures the user's ear scan, verification can be performed using efficient binary operations, such as XOR.

\smallskip
\noindent{\underline{\circled{2} Raw biometric data transmission during authentication.}}
Raw biometric data is immutable and sensitive; if compromised, it can be reused indefinitely for unauthorized access \cite{appleBiometricSecurity}. To mitigate this risk, systems must avoid exposing such data beyond the user's control. For instance, Apple’s Touch ID processes fingerprints within a hardware-isolated Secure Enclave \cite{AppleTouchID, appleSecureEnclave}. In contrast, wireless earbuds rely on Bluetooth, making them more vulnerable during authentication. While enrollment may occur in trusted settings, authentication often takes place in less secure environments, where adversaries may exploit Bluetooth attacking tools to capture and replay audio signals, such as BlueSpy \cite{tarlogicBlueSpy}. This motivates our approach: to extract the ECS key solely on the earbuds and perform verification on the mobile device.

\smallskip
\noindent \textbf{Motivation: extract on earbuds, verify on mobile.} \\
Towards our motivation, we propose \textbf{EarID}, a key extraction protocol that encodes distinctive biometric features into a fixed-length binary key. Instead of training classifiers during enrollment, EarID identifies user-specific uniqueness by comparing the user’s feature distribution against population-level statistics. The most distinctive features are binarized into keys compatible with privacy-preserving schemes such as fuzzy commitment~\cite{fuzzy_commitment} or FIDO2~\cite{fido2}. During authentication, only the binary keys are transmitted over Bluetooth. Because raw biometric data is never shared, intercepted communication cannot reveal any ear canal information. Moreover, EarID avoids the overhead of model training, enabling efficient on-earbud computation and responsive authentication.

\smallskip
\noindent \textbf{Contributions.}\

\noindent \textbf{A. Learning-free ear canal key extraction.}
To address challenge \circled{1}, we propose a learning-free ECS biometric key extraction algorithm to generate unique keys for individual users. By applying denoising and physically-meaningful transformation, we leverage biometric information \cite{BioInfoRE} for feature selection and utilize the random matrix method \cite{BioHashing} for binarization. Such a process speeds up about 3× to 90× enrollment time, comparing to different ML classifiers training time on mobile. Meanwhile, it enables 160~ms processing time on earbuds MCU\footnote{Unless otherwise specified, the EarID performance is reported based on BCH(255,123,19).}, which is negligible to user authentication experience.

\noindent \textbf{B. Privacy-preserving earbuds-to-mobile authentication.}
To address challenge \circled{2}, we eliminate raw data and explicit key transmission during the authentication phase, when the wireless environment may be untrusted (e.g., subject to interception or eavesdropping). Specifically, the binary key enables the use of Fuzzy Commitment \cite{fuzzy_commitment}, a privacy-preserving biometric authentication scheme well-suited for secure verification between wireless earbuds and mobile devices—without exposing raw biometric data.

\noindent \textbf{C. EarID testbed and evaluation.}
We develop a complete hardware-software testbed for ear acoustic scanning, data collection, and processing. Using real-world collected data, we evaluate EarID in terms of accuracy, efficiency, robustness, and security under various simulated and emulated conditions.

\section{Background \label{sec:background}}
\subsection{Biometric Authentication System}
\noindent \textbf{Framework.}
A standard biometric authentication system consists of four main modules: a biometric data acquisition device, a biometric feature extractor, a biometric database, and an authentication decision-making module, such as a classifier or matcher. Specifically, the data acquisition step involves using a biometric sensor (e.g., a camera for face recognition, a microphone for voice authentication, or an earbud's speaker and in-ear microphone for ECS) to collect physiological or behavioral signals from the legitimate user, known as the \textit{biometric sample}. The \textit{biometric sample} is then extracted into the \textit{biometric feature}, which contains discriminative information specific to the user and is resistant to unrelated factors. A \textit{biometric template} stores the legitimate user's information in a secure database for further authentication. This template can consist of (un)transformed features, feature distributions, or machine learning models derived from the features. Finally, the decision-making is done by a classifier or a matcher that verifies whether the current attempt belongs to the claimed user. 

\smallskip
\noindent \textbf{Enrollment and Authentication.}
In the enrollment phase, the system captures the raw \textit{biometric samples} from the user, extracts \textit{biometric features}, and stores the \textit{biometric template} in the database. In the authentication phase, the system will capture the new \textit{biometric sample}, follow the same data processing, and make decisions based on the stored \textit{biometric template}.

% \smallskip
% \noindent \textbf{Local vs. Remote Biometric Authentication.}
% Local biometric authentication involves verifying a user's identity directly on the device. For example, unlocking a smartphone with a fingerprint or accessing a laptop with facial recognition. It is fast and does not rely on network connectivity but may be less secure if the device is compromised.
% Remote biometric authentication, on the other hand, verifies identity by communicating with an external server or service. This method is commonly used for online accounts and remote access, such as logging into a secure website. It can leverage centralized security measures, offering potentially higher security but relies on network connectivity and can be slower due to the need for data transmission.

\subsection{Ear Canal Scanning}
\noindent \textbf{Ear Canal Scanning Front-End.}
Many ECS authentication papers \cite{2019earecho, 2021eardynamic, 2021headfi, 2023earace} adopt a radar/sonar-style in-ear acoustic sensing approach. A speaker emits a wide-band waveform into the ear canal, and the in-ear microphone records the reflected signal, as shown in Fig.~\ref{fig:ear-scan}. The acoustic channel is estimated via cross-correlation:
\begin{equation}
\label{eq:1-channel-estimation}
    h(n) = \sum_{t=0}^{\infty} s(n-t)r(t),
\end{equation}
where \( h(n) \) is the target in-ear impulse response at sample index \( n \), \( s(t) \) is the transmitting waveform at time \( t \), and \( r(t) \) is the reflected signal captured by the in-ear microphone. Nevertheless, the magnitude transfer function, which is the magnitude of the Fourier Transform of \( h(n) \), is employed by most papers due to the smaller intra-user variation by removing the noisy phase information. Prior work~\cite{2018inaudibleEar,2023earace,2019earecho,2021eardynamic} has also explored waveform design, re-insertion effects, device response, and canal deformation.

\smallskip
\noindent \textbf{Decision Making Back-End.} 
Most ECS authentication studies rely on machine learning classifiers such as Support Vector Machine (SVM)~\cite{2019earecho,2021headfi,2022bilateral}, Linear Discriminant Analysis (LDA)~\cite{2018inaudibleEar}, and ensemble learning methods. In contrast, template matching, a commonly used back-end in fingerprint and iris recognition, has not yet been adopted in ECS systems. Template matching operates on fixed-length feature vectors without model training, making it lightweight and compatible with privacy-preserving authentication frameworks. However, it requires compact, stable, and user-specific representation from the raw biometric samples, which has not been explored yet for the ear scan data. In this work, we, for the first time, propose the ECS key extraction protocol that can practically run on wireless earbuds.

\begin{figure*}[t]
    \centering
    \includegraphics[width=\textwidth]{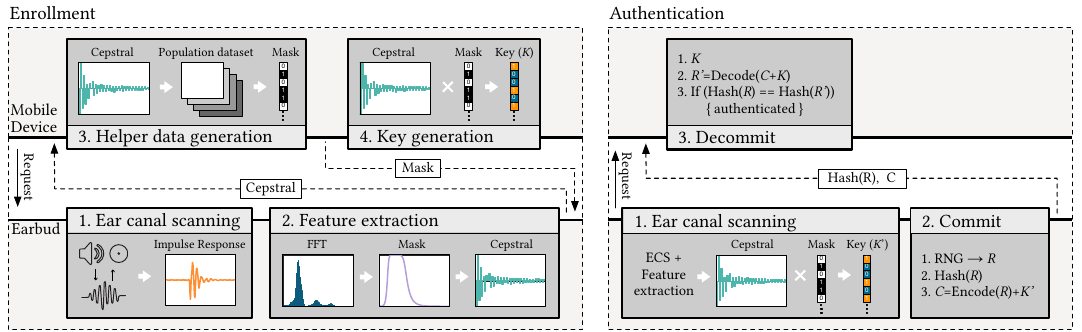}
    \caption{EarID authentication protocol for enrollment and authentication stage. Enrollment: earbuds locally run (1) ear canal scanning (ECS), (2) cepstral feature extraction and data transmission; and then mobile devices runs (3) helper data generation (send to earbuds) for (4) key extraction (stored at mobile devices). Authentication: wireless earbuds run (1) on-device key extraction based on new ECS data, (2) commitment generation and transmission; and then mobile devices runs (3) decommitment and verification.}
    \label{fig:protocol}
\end{figure*}

\section{EarID System Design \label{sec:system-overview}}
\subsection{System and Threat model}
\noindent \textbf{System Model.}   
The goal of the proposed \textit{EarID} system is to enable ECS-based biometric authentication that supports a hands-free, listen-to-authenticate experience. The system is designed for execution on commercial wireless earbuds, allowing ECS biometric key extraction directly on the device. We consider a typical use case in which a user wears wireless earbuds paired with a personal mobile device, such as a smartphone. The ear canal features serve as the user's biometric trait for authentication. To enable this, we assume that the wireless earbuds are equipped with in-ear microphones and operated by an embedded microcontroller (MCU) and a digital signal processing (DSP) unit capable of audio computation \cite{nxpEarbuds}. A Bluetooth module allows communication between the earbuds and the paired mobile device.
The mobile device, owned by the user, is equipped with a general-purpose processor and sufficient storage. It is responsible for storing relevant authentication data, including population-level feature statistics and user-specific keys. We assume this sensitive data is stored in a protected location such as the Secure Enclave \cite{appleSecureEnclave}, which cannot be accessed or modified by unauthorized parties.

To use EarID, the user must wear the earbuds firmly inserted into the ear canal and ensure the mobile phone is within Bluetooth communication range. We assume the enrollment process takes place in a secure environment where the user has physical control of both devices and a trusted wireless channel. During enrollment, the mobile device receives raw ECS data from the earbuds to generate and store the user's reference key. The mobile also provides the corresponding helper data to the earbuds.
For subsequent authentication, the same earbuds and mobile device are used. The earbuds collect a fresh ECS sample, regenerate the biometric key locally, and transmit commitment data to the mobile device. The mobile verifies the user's identity based on the received data and the stored reference.

%
% During the initial enrollment, the EarID scheme performs an ECS scan, extracts a binary key, stores this key securely on the mobile device, and sends associated helper data back to the earbuds. During subsequent authentication, the earbuds use this helper data to regenerate a matching key from the same user’s ECS scan. Instead of transmitting directly, EarID adopts a fuzzy commitment scheme: the earbuds encrypt a random binary sequence using the regenerated key to form a commitment. They then transmit this commitment along with a hash of the random sequence to the mobile device. The mobile verifies the user's identity by recovering the original sequence using the stored key and checking whether its hash matches the one received from the earbuds. If the hashes match, the user is successfully authenticated.
%

\smallskip
\noindent \textbf{Threat Model.}  
We consider an adversary whose goal is to impersonate a legitimate user and gain unauthorized access to the user’s mobile device, knowing that the user relies on wireless earbuds for ECS-based authentication. Although wireless earbuds and mobile devices are typically personal belongings used exclusively by the owner, we also account for cases where these devices may be temporarily accessed by close individuals, such as co-workers or curious family members. We further consider more sophisticated scenarios in which the devices are stolen by the adversary.

We categorize the threat model into two main scenarios, namely with physical access to devices or without physical access.
In the first case, the attacker has direct access to both the victim’s mobile device and wireless earbuds. The attacker may attempt the following:
\begin{itemize}
    \item \textit{Passive attack.} The attacker inserts the victim’s earbuds into their own ear and attempts authentication using their own ECS response.
    \item \textit{Synthetic ECS attack.} The attacker inserts the victim’s earbuds into a silicon ear model in order to spoof the ECS response.
    \item \textit{Universal ECS attack.} A more sophisticated synthetic attack, where the attacker constructs a silicon ear model designed to emulate common ECS patterns extracted from public datasets.
\end{itemize}
In the second case, the attacker does not possess the devices but remains within Bluetooth range to launch an attack, for example using tools like BlueSpy~\cite{tarlogicBlueSpy}. A common strategy is:
\begin{itemize}
    \item \textit{Key guessing attack.} The attacker attempts to authenticate by transmitting a guessed or self-generated ECS key via a compromised or previously paired Bluetooth connection.
\end{itemize}
It is important to note that under proximity-based conditions, the adversary could also attempt biometric stealing if raw ECS data were transmitted over the Bluetooth channel. However, our design eliminates this threat by avoiding the transmission of raw biometric information altogether, thereby rendering such attacks infeasible.

\subsection{Proposed Protocol}

\noindent \textbf{Overview.}  
EarID is a biometric authentication protocol based on ear canal scanning key extraction, designed for seamless integration with commercial wireless earbuds and mobile devices. Unlike conventional approaches that rely on classifiers or transmit raw biometric data during authentication, EarID generates a user-specific binary key directly on the earbuds, enabling secure and private authentication without exposing sensitive information. The system consists of two core components: the wireless earbuds, which capture and process acoustic signals from the user’s ear canal, and the mobile device, which verifies user identity based on a previously stored key. By avoiding machine learning models and minimizing computational burden, EarID delivers a lightweight, privacy-preserving, and user-friendly authentication experience.

\smallskip
\noindent \textbf{Wireless Earbuds.}  
The wireless earbuds serve as the sensing and processing front end of the EarID system. When worn by the user, the earbuds utilize their built-in audio unit to perform ECS by emitting acoustic energy and capturing the impulse response, as defined in Eq.~(\ref{eq:1-channel-estimation}). The captured ECS signal is then processed locally to extract relevant features. With the assistance of helper data (i.e., user-specific mask) received during enrollment, the earbuds can reliably regenerate the user-specific key during authentications.

\smallskip
\noindent \textbf{Mobile Device.}  
The mobile device plays two primary roles within the EarID framework. First, it has access to population-level ECS statistics and assists in user enrollment by generating and transmitting helper data specific to the user’s wireless earbuds. Second, it functions as the identity verifier: upon receiving authentication data from the earbuds, the mobile device performs verification using the fuzzy commitment protocol, comparing the received commitment against the stored credential to determine authenticity.

\subsection{Authentication Process} 
\noindent \textbf{Enrollment Phase.}  
The enrollment process begins when the user wears the wireless earbuds and pairs them with the mobile device. As shown in Fig.~\ref{fig:protocol}, the process is initiated by the mobile device.  
\circled{1}: The earbuds perform ear canal scanning to capture the ear canal impulse response.  
\circled{2}: The earbuds extract features from the scanned signal using cepstral analysis and transmit the resulting cepstral features to the mobile device.  
\circled{3}: The mobile device analyzes the user’s ECS feature uniqueness based on population-level statistics and generates corresponding helper data, including a feature mask and binarization parameters.  
\circled{4}: The mobile generates and securely stores the user’s key, then sends the helper data back to the earbuds for future authentications.

\smallskip
\noindent \textbf{Authentication Phase.}  
When the user wears earbuds to request authentication, such as unlocking the device or authorizing a transaction, the following steps occur:  
\circled{1}: The earbuds perform ECS and extract features, then use the stored helper data to generate a new biometric key.  
\circled{2}: Following the fuzzy commitment protocol, the earbuds encrypt a random secret (e.g., binary string) using the newly generated key and transmit the resulting commitment and a hash of the secret to the mobile device.  
\circled{3}: The mobile device receives and decodes the random secret with stored key. If the recovered hash matches the transmitted hash, the user is successfully authenticated.
All steps above are illustrated in Fig.~\ref{fig:protocol}.

\begin{figure*}
    \centering
    \begin{minipage}[t]{0.48\linewidth}
        \centering
        \includegraphics[width=\linewidth]{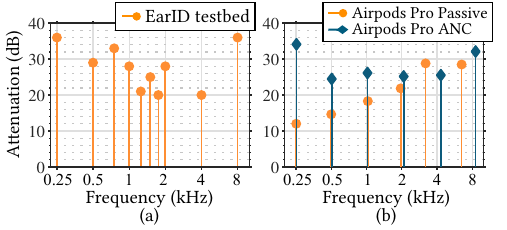}
        \caption{Impact of passive and active noise cancellation (ANC) for (a) EarID testbed and (b) Apple Airpods Pro \cite{airpods-noise-cancelling}.}
        \label{fig:attenuation}
    \end{minipage}
    \hfill
    \begin{minipage}[t]{0.48\linewidth}
        \centering
        \includegraphics[width=\linewidth]{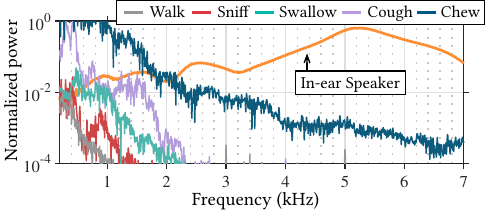}
        \caption{Impact of occlusion effect amplified body sound, interfering acoustic sensing at low frequencies ($<$2~kHz)}
        \label{fig:in-body-sound}
    \end{minipage}
\end{figure*}
\section{Learning-Free ECS Key Extraction\label{sec:client-design}}
\subsection{Feature Extraction} 
\smallskip
\noindent \textbf{ECS Denoising.}
ECS noise can originate from external environments or in-body sounds. External noise, such as car horns, birds singing, and background music, is less significant for the in-ear microphones. Since the earbuds are inserted firmly into the ear canal for better experiences, the ear tips provide significant attenuation, up to 40 dB as shown in Fig.~\ref{fig:attenuation}. 
On the other hand, when the ear canal is occluded, it also amplifies low-frequency in-body sounds by up to 40-dB~\cite{ear-occlusion-effect}, including sounds from breathing, walking, and eating activities. These sounds can have power levels comparable to or even exceeding the ear canal response below 2~kHz.

In Fig.~\ref{fig:in-body-sound}, we observe that noises mostly interfere with the low-frequency band (below 2~kHz) due to their impact on the occlusion effect. To mitigate this, we apply a bandpass filter to exclude the vulnerable low-frequency band below 2~kHz for user authentication applications. As a result, we set the low cut-off frequency at 2~kHz and recommend setting the high cut-off frequency within the device frequency range, e.g., 8~kHz, which offers flat frequency response. Next, we perform power normalization to ensure consistent signal strength and improve the robustness of the biometric features against variations. In addition, data aggregation by averaging multiple scanning results can mitigate the random variation caused by the dynamic deformation, especially for short waveform duration. Therefore, the normalized band-pass FFT feature is given by
\begin{equation}
\label{eq:3-FFT-norm}
    \left|H^{\text{norm}}(f)\right| = \frac{\sum_{k=1}^{K} \left| H^{\text{est}}_{k}(f) F^{\text{bp}}(f) \right|}
    {\sqrt{\sum_{f_l}^{f_h} \sum_{k=1}^{K} \left| H^{\text{est}}_{k}(f) F^{\text{bp}}(f) \right|^2 }},
\end{equation}
where \( |H^{\text{norm}}(f)| \) is the normalized magnitude of the frequency response,  \( H^{\text{est}}_{k}(f) \) is the ear canal frequency response measured in \(k\)-th trial, \( F^{\text{bp}}(f) \) represents the bandpass filter, \( K \) is the total number of trials, and \( f_l \) and \( f_h \) denote the lower and upper bounds of the frequency range.

\begin{figure}[t]
    \centering
    \includegraphics[width=\linewidth]{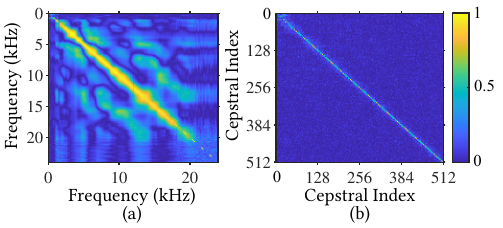}
     \caption{Cross-correlation of (a) FFT features (strong correlation) and (b) cepstral features (independent).}
    \label{fig:fft-cepstrum-compare}
\end{figure}

\smallskip
\noindent \textbf{Cepstral Analysis.}
The FFT-normalized feature in Eq.~(\ref{eq:3-FFT-norm}) can be highly correlated and redundant, i.e., inconvenient to add or remove one dimension based on its importance. We aim to compress this ECS representation and preserve the most useful information. Therefore, we apply cepstral analysis to the FFT feature vector, which reveals periodic structures that indicate echoes and reverberations in the quefrency domain \cite{signal-and-systems}. This process generates a new cepstrum feature vector whose entries are uncorrelated and can be interpreted as unique multipath reflections within the ear canal. The cepstrum feature is defined as:
\begin{equation}
    c(n) = \mathcal{F}^{-1} \left\{ \log \left|H^{\text{norm}}(f)\right| \right\},
\end{equation}
where \(c(n)\) denotes the cepstrum and \( \mathcal{F}^{-1} \) is the inverse Fourier Transform with ceptrum index \(n\).  Because the FFT feature is a real-value vector, we use the inverse Discrete Cosine Transform (IDCT) instead. Each entry of the transformed feature \(c(n)\) represents a unique multipath strength from the ear canal. Given the ear canal length and the speed of sound, we can safely remove cepstral components outside the reasonable time-of-flight range (e.g., 10 ms). This step is called time-of-flight liftering \cite{signal-and-systems}, which reduces feature dimension and removes irrelevant noises. As a result, the cross-correlation among features in a user's feature vector can be reduced, as shown in Fig.~\ref{fig:fft-cepstrum-compare}.

\subsection{Key Generation}
\noindent \textbf{Biometric Information.}
EarID must identify the informative multipath from a user's enrollment cepstrum features. This ``information" is considered high if a feature exhibits low intra-subject variation and high inter-subject variation. Using handwriting recognition as an analogy, Alice's distinct handwriting style makes her writing easily recognizable, and her handwriting remains consistent across different instances. To this end, we utilize the \textit{biometric information} (BI) defined by Adler et al. \cite{BioInfoRE} as a metric, which is the distribution divergence between the feature distribution from the user enrollment data and that from the population. We modify the original definition using the Rényi divergence as:
\begin{equation}
\label{eq:bioinfo}
D_{\alpha}(P \| Q) = \frac{1}{\alpha - 1} \log_2 \left( \sum_{x \in \mathcal{X}} P(x)^\alpha Q(x)^{1-\alpha} \right),
\end{equation}
where \( P(x) \) represents the probability distribution of the user's feature, \( Q(x) \) represents the probability distribution of the population's feature, and \( \alpha \) is the order of the Rényi divergence. This modification allows us to focus on different aspects of distribution divergence by tuning \( \alpha \). The empirical study shows that when \( \alpha\rightarrow0\), the EarID system achieves the best overall performance.

\smallskip
\noindent \textbf{Feature Masking.}
It is straightforward to generate a feature mask, select cepstrums with relatively high BI, and discard those with low BI that may lead to errors. However, setting a hard BI threshold for all users is not desirable. For instance, if a user has uniformly high BI cepstrums, only a subset of the highest BI features is sufficient. Conversely, for a user with generally low BI, the system should still attempt to recognize them using their most informative cepstrums. Therefore, we propose the masking approach based on each user's BI across cepstrum features by automatic thresholding using the Otsu's method \cite{ostu-method}. The resultant user-specific binary feature mask is then applied to the cepstrum feature, where `1' means the high-BI cepstrum to retain and `0' means the low-BI cepstrum to discard.

\smallskip
\noindent \textbf{Feature Binarization.}
Given the selected features with high biometric information, the final step is to generate a fixed-length binary key by transforming the feature vector from continuous space into a binary string. Since features are standardized using population-level mean and variance, i.e. \(\tilde{c}=(c-\bar{c})/\text{var}(c) \), each dimension has approximately a 50\% chance of being positive or negative, making sign-based binarization a natural choice. However, using only the sign discards magnitude information and limits the key length to the number of selected features.

\begin{figure}[t]
    \centering
    \includegraphics[width=1\linewidth]{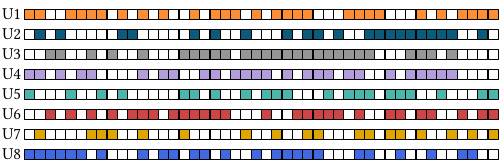}
    \caption{The binary ECS codes transformed from frequency responses from 8~users, preserving the uniqueness among users. \label{fig:user_code}}
\end{figure}

To address this, we adopt a random projection approach, also known as BioHashing \cite{BioHashing}. A random matrix \( \mathbf{R} \in \mathbb{R}^{L \times d} \) with entries drawn from a standard normal distribution is multiplied with the standardized feature vector \( \mathbf{x} \in \mathbb{R}^{d \times 1} \), producing a projected vector \(\mathbf{y} = \mathbf{R} \cdot \mathbf{x}\).
The final binary key \( \mathbf{K} \in \mathbb{B}^{L \times 1} \) is then obtained by applying a sign function to each element:
\begin{equation}
\mathbf{K}[i] = 
\begin{cases}
1, & \text{if } \mathbf{y}[i] \geq 0, \\
0 ,& \text{otherwise},
\end{cases}
\end{equation}
and the visualization of the derived key is shown in Fig.~\ref{fig:user_code}. In the figure, eight different users all have their unique key extracted from the ECS data, where similar number of colored boxes (bit `1') and white boxes (bit `0') can be observed.

\begin{figure*}[t]
    \centering
    \includegraphics[width=0.99\linewidth]{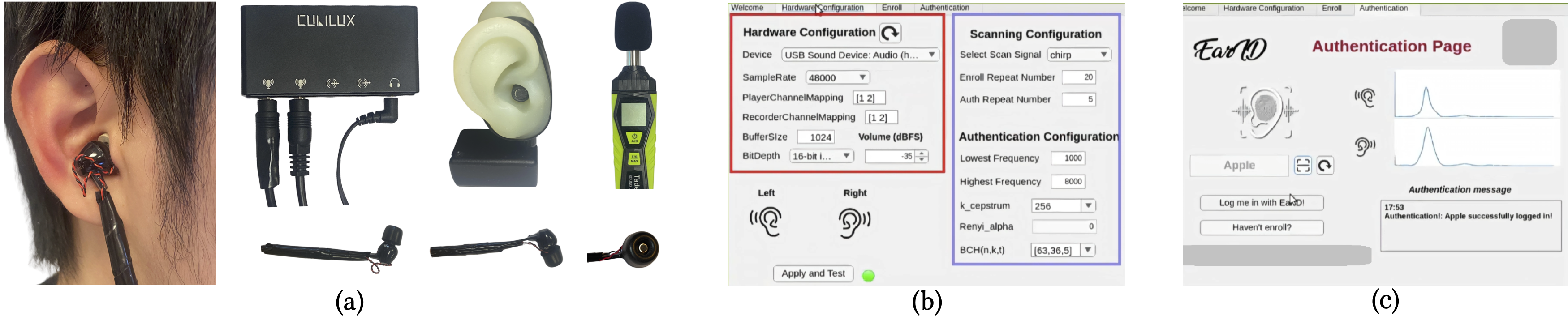}
    \caption{(a) EarID hardware platform (b) EarID software configuration page and (c) EarID authentication page (dummy user)}
    \label{fig:UI}
    \vspace{-5mm}
\end{figure*}

\section{Earbuds-to-Mobile Authentication\label{sec:client-design}}
Since the biometric key representing the legitimate user's identity has already been generated, we now describe the authentication protocol based on these keys. As introduced in previous sections, the mobile device securely stores the enrolled key \(\mathbf{K}^\text{enroll}\), while the wireless earbuds are enabled to generate a fresh key \(\mathbf{K}^\text{auth}\) from ECS during each authentication attempt. While this design avoids transmitting raw biometric data over potentially insecure channels, the challenge is shifted to: how can the mobile device verify that the key generated on the earbuds matches the enrolled one on the mobile side?

To address this, we adopt the fuzzy commitment scheme, which allows verifying key similarity from two devices (earbuds and mobile). During each authentication, the earbuds use a random number generator (RNG) to produce a random (binary) secret \(\mathbf{R} \in \mathbb{B}^{T \times 1}\), where \(T < L\). This random sequence provides \(\log_2(T)\) bits of security. The hash of random sequence \(\text{Hash}(\mathbf{R})\) is transmitted to the mobile device for verification. Besides, wireless earbuds use an error-correcting code (ECC), such as BCH or Reed-Solomon, to encode \(\mathbf{R}\) into a codeword of length \(L\), denoted \(\text{Enc}(\mathbf{R})\). The encoded sequence is then combined with newly generated key:
\[
\mathbf{C} = \text{Enc}(\mathbf{R}) \oplus \mathbf{K}^\text{auth},
\]
where \(\mathbf{C}\) is the commitment transmitted to the mobile device. The ECC enables correction of up to \(t\) bit errors, compensating for noise and variability in ECS-based key extraction.

Upon receiving the commitment \(\mathbf{C}\) and hash \(\text{Hash}(\mathbf{R})\), the mobile device uses the stored key \(\mathbf{K}^\text{enroll}\) to attempt recovery of the original message:
\[
\mathbf{R}' = \text{Dec}(\mathbf{C} \oplus \mathbf{K}^\text{enroll}),
\]
where \(\text{Dec}(\cdot)\) denotes ECC decoding. The mobile then computes \(\text{Hash}(\mathbf{R}')\) and compares it to the received \(\text{Hash}(\mathbf{R})\). If the two match, it implies that the key generated during authentication is close to the enrolled key within the ECC correction capability, i.e.,
\[
\text{dist}(\mathbf{K}^\text{auth}, \mathbf{K}^\text{enroll}) \leq t.
\]

The primary benefit of adopting the fuzzy commitment scheme is that it eliminates the need to transmit raw biometric data, thereby mitigating the risk of biometric theft during communication. Instead of directly revealing the biometric key or ear canal impulse response, the scheme verifies the user's identity through indirect comparison. The choice of ECC in the commitment process reflects a trade-off between security and tolerance to intra-user variability; specifically, it determines the number of protected bits (entropy) and the system’s ability to correct bit-level errors. In this work, we adopt BCH codes for the EarID fuzzy commitment system. For example, a BCH(255,123,19) code implies a key and codeword length of \(L = 255\), provides \(T = 123\) bits of protection, and corrects up to \(t = 19\) bits of error.

\section{Evaluation}
In the following section, we will evaluate the proposed system and use the acronyms explained in Table~\ref{tab:acro}.

\begin{table}[ht]
\centering
\caption{List of Acronyms}\label{tab:acro}
\small
\begin{tabularx}{\linewidth}{@{} l Y  l Y @{}}
\toprule
\textbf{Acronym} & \textbf{Description} & \textbf{Acronym} & \textbf{Description}\\
\midrule
ECS & Ear Canal Scanning            & SVM & Support Vector Machine \\
BER & Bit Error Rate                & MLP & Multi-Layer Perceptron \\
FRR & False Rejection Rate          & RF  & Random Forest \\
FAR & False Acceptance Rate         & ASR & Attack Success Rate \\
EER & Equal Error Rate              & RNG & Random Number Generator \\
BAC & Balanced Accuracy             & ECC & Error-Correcting Code \\
BCH & Bose–Chaudhuri–Hocquenghem Code &   & \\
\bottomrule
\end{tabularx}
\end{table}

\subsection{Data Collection \label{sec:data-collection}} 
\noindent \textbf{EarID Hardware Platform.}
While commercial wireless earbuds do not currently provide access to in-ear microphone APIs, we developed a custom hardware setup to emulate their functionality. Specifically, we embedded mini microphones (CMC-3015-44L100~\cite{in-ear-mic-sensor}) into the foam tips of off-the-shelf (COTS) earbuds (Sony MDREX15AP~\cite{SonyMDREX15AP}). These microphones, positioned toward the ear canal, worked in tandem with the earbuds' built-in speakers to perform ear canal scanning, functionally replicating the ECS capability of wireless earbuds. Both the speaker and microphone were connected to a full-duplex USB sound card, which served as a proxy for the DSP typically found in wireless earbuds. The sound card was interfaced with a laptop for data acquisition and analysis. The complete hardware platform is illustrated in Fig.~\ref{fig:UI}(a). In addition, we utilized a silicon ear model to simulate the \textit{Synthetic ECS attack} scenario, and a Sound Pressure Level (SPL) meter was employed to monitor ambient noise levels during experiments. 

\begin{table}[h]
\caption{Acoustic and EarID System Parameters}
\label{tab:audio_params}
\resizebox{\linewidth}{!}{%
\begin{tabular}{ll|ll}
\toprule
\textbf{Acoustic Parameter} & \textbf{Value} & \textbf{Acoustic Parameter} & \textbf{Value} \\
\midrule
Volume & User-adjustable & Dataset Samples  & 44$\times$40=1760 \\
Sampling Frequency & 48 kHz & Scans/Enroll in Wild & 8 \\
MLS Duration & 1 second & Scans/Auth in Wild & 2 \\
Chirp Frequency & 20 Hz - 20 kHz & Quiet Room SPL & 30-40 dBA \\
Chirp Duration & 1 second & Indoor/Outdoor Noise & $\sim$60/70 dBA \\
\bottomrule
\end{tabular}%
}
\end{table}
\begin{table*}[]
\centering
\caption{Performances of EarID Authentication \label{tab:earid-result}}
\begin{tabular}{lllllcllll}
\toprule
\textbf{EarID w. ECC}          & \multicolumn{4}{c}{\textbf{Accuracy}} & \multicolumn{1}{c}{\textbf{Robustness}} & \multicolumn{4}{c}{\textbf{Security}}  \\
\cmidrule(lr){2-5} \cmidrule(lr){6-6} \cmidrule(lr){7-10}
\textbf{Metrics} (\%)             & \textbf{EER} $\downarrow$  & \textbf{FAR} $\downarrow$   & \textbf{FRR} $\downarrow$  & \textbf{BAC} $\uparrow$  & \textbf{FT-Acc. Rate} $\downarrow$                  & \textbf{P-ASR} $\downarrow$ & \textbf{S-ASR} $\downarrow$  & \textbf{U-ASR} $\downarrow$  & \textbf{K-ASR} $\downarrow$  \\
\cmidrule{1-1}

BCH(127,64,10)  & 1.2   & 1.0   & 2.2   & 98.4  &          0.1            &      1.0       &      0.8        &     0.2         &         0.0     \\
BCH(255,123,19) & 0.7   & 0.6   & 2.0   & 98.7  &           0.0            &       0.6       &      0.8        &      0.1        &      0.0        \\
BCH(511,241,30) & 1.0   & 1.0   & 2.3   & 98.4  &            0.0           &      1.0        &       0.7       &     0.0         &        0.0      \\
\bottomrule
\end{tabular}
\end{table*}

\noindent \textbf{EarID Software Platform.}
To facilitate the testing process, we developed a software platform for data collection and evaluation. The configuration page, shown in Fig.~\ref{fig:UI}(b), allowed the developer to adjust system parameters, including hardware configuration (in red box) and authentication configuration (in purple box). During enrollment, users simply registered their identifier (non-sensitive last 4 digit of their phone number) and captured their ECS signal for key extraction. The extracted key was then stored. For authentication, as depicted in Fig.~\ref{fig:UI}(c), users typed in their identifier and performed quick ECS scans. The system provided real-time visual feedback to help users monitor signal quality and displayed final authentication results on the interface. With the user consent, we collected the enrollment and authentication data for algorithm development.

\smallskip
\noindent \textbf{IRB-Approved Subject Recruitment.}
This experiment, approved by IRB protocols, involved recruiting 44 participants from a university campus, ensuring diversity in gender, race, age, and roles (students, faculty, and staff). Our user pool surpasses the median size (26) of similar studies \cite{2018inaudibleEar,2019earecho,2021eardynamic,2021headfi,2022bilateral,2023earace} and is second only to the largest pool (52) in \cite{2022bilateral}. For cross-day evaluation, due to scheduling challenges, data was collected from 8 participants.

\smallskip
\noindent \textbf{Ear Scanning and Recording.}
Human subject data collection for ear canal scanning was conducted in a quiet indoor environment with ambient noise levels ranging from 30 to 40 dBA. Participants were instructed to insert the earbuds until their ears were fully occluded, remain silent, and behave naturally throughout the recording process using our platform in Fig.~\ref{fig:UI}. Each data collection session consisted of 20 repetitions of ECS, and each user completed two sessions using both Maximum-Length Sequence (MLS) and chirp signals. The specific parameters for these excitation signals are summarized in Table~\ref{tab:audio_params}.

\smallskip
\noindent \textbf{Robustness Experiments.}
For robustness evaluations, only chirp signals were used in the false triggering, cross-day and cross-session tests. To evaluate false-triggering, which are scenarios referred to as the ``not-in-use'' condition, data were collected while the earbuds were held in hand, placed on various surfaces, or covered by cloth. Cross-day testing involved five authentication attempts per day over five consecutive days, while cross-session testing consisted of one enrollment session followed by nine authentication sessions conducted within a short time window. A silicon ear model was also employed as a baseline for comparison with human users and to simulate passive attacks in the cross-session evaluation.

\smallskip
\noindent \textbf{Evaluation Parameters.}
To evaluate the performance of EarID, we focus on authentication accuracy, efficiency, robustness against unintended false triggering, and security against attacks. The default setup of the experiment is as follows. The number of FFT points is set to 1024, and the number of cepstrum coefficients after liftering is 256. The bandpass denoising filter is designed to allow frequencies between 2 kHz and 8 kHz. The order of the Rényi divergence \(\alpha\) is 0.

\begin{figure*}[t]
    \centering
    \includegraphics[width=\textwidth]{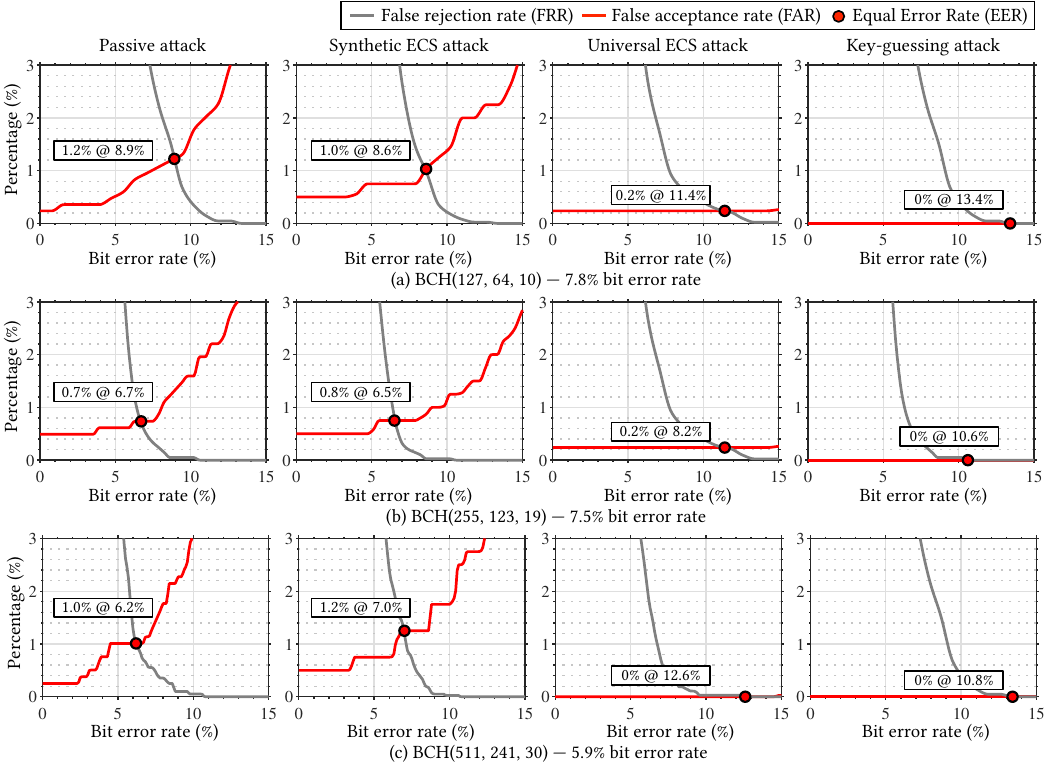}
    \caption{The false rejection rate (FRR, in gray) and false acceptance rate (FAR, in red) plots against the bit error rate (BER) between enrolled and authentication keys. The intersection of FRR and FAR is the point achieving equal error rate (EER), where the corresponding BER is the ideal error correction threshold. The actual error correction codes (ECC) are (a) BCH(127,~64,~10), (b) BCH(255,~123,~19), and (c) BCH(511,~241,~30).}
    \label{fig:bit-error-eer}
\end{figure*}

\subsection{EarID Accuracy Evaluation} 
\noindent \textbf{Bit Error Interpretation.} 
To evaluate the performance of key extraction, we first directly compare the enrolled key stored on the mobile device with fresh keys generated during authentication attempts to gain insight. A low bit error rate (BER) is expected for legitimate authentication, indicating strong key consistency for the same user. In contrast, a high BER is desirable when comparing keys from different users (i.e., passive attackers), reflecting good user discrimination. We evaluate three key lengths \(L\): 127, 255, and 511. Results are shown in ``Passive attack" in Fig.~\ref{fig:bit-error-eer}, where the gray and red lines represent BERs for legitimate users and passive attackers, respectively. In the figure, above 97\% percentage legitimate user only have approximately 5\% BER. When BER increases to more than 10\%, the percentages of the legitimate user gradually drops to 0. In contrast, red line shows that the percentage of attackers is extremely low at low BER region and gradually increases above 15\% BER for passive attack (human intruders). Since line plots in Fig.~\ref{fig:bit-error-eer} reflect the changes of false rejection rate (FRR) and false acceptance rate (FAR), their intersection (red circle) refers to the optimal threshold, achieving equal error rate (EER).

\begin{table}[h]
\centering
\caption{Accuracy of ML Classifier Baselines \label{tab:ML-baseline}}
\begin{tabular}{lllll}
\toprule
\textbf{ML Classifiers} & \multicolumn{4}{c}{\textbf{Accuracy}} \\ \cmidrule(lr){2-5}
\textbf{Metrics} (\%)        & \textbf{EER} $\downarrow$   & \textbf{FAR} $\downarrow$  & \textbf{FRR} $\downarrow$  & \textbf{BAC} $\uparrow$   \\ \cmidrule{1-1}
SVM            & 0.9   & 1.3   & 0.7  & 99.0  \\
Random Forest  & 0.7   & 2.9   & 0.5  & 98.3  \\
MLP            & 1.0   & 1.2   & 0.9  & 99.0  \\
Adaboost       & 2.0   & 1.7   & 2.3  & 98.0  \\ \bottomrule
\end{tabular}
\end{table}

\smallskip
\noindent\textbf{Authentication Accuracy and ECC.}
In practice, the EarID system does not have the bit error analysis data to make decision. Instead, the decision is made by a selected ECC. If the bit error goes beyond certain error correction capacity, the decommitment results will never match (see Fig.~\ref{fig:protocol}), leading to reject the attempt. Therefore, the ECC should be selected to provide sufficient key length as well as a suitable error correction ability near the exact EER point.
For example, with a key length \(L=127\), the corresponding EER corresponds to tolerating 8.9\% BER (about 11 bits). Using BCH(127,64,10), which can correct 10 bit errors, it will achieve a FRR of 2.2\% and FAR of 1.0\%, close to the ideal EER. Similarly, performance with BCH(255,123,19) and BCH(511,241,30) is reported in Table.~\ref{tab:earid-result}. We also report the Balanced Accuracy (BAC) defined as \(\text{BAC}=(2-\text{FAR}-\text{FRR})/2\) in the table.

\smallskip
\noindent \textbf{Classifier Baseline Configurations.} We evaluate EarID against four baseline machine learning classifiers: Random Forest (RF), Support Vector Machine (SVM), Multi-Layer Perceptron (MLP), and AdaBoost. All models are implemented using scikit-learn~\cite{scikit} and tuned to achieve optimal performance. The RF model uses 50 decision tree estimators with a maximum depth of 10. The
SVM adopts a linear kernel with a regularization parameter \(C=50\). The MLP consists of two hidden layers with 16 ReLU units each and is trained using the Adam optimizer for up to 1000 iterations. AdaBoost integrates 10 weak decision trees, each with a maximum depth of 10.

\smallskip
\noindent\textbf{Comparison with Classifiers.}
To train the classifiers, we follow the protocol described in \cite{2019earecho}. The target user’s enrollment data is combined with data from 30 randomly selected users (also referred to as built-in users) to train the model. An additional 10 users are designated as passive attackers. For evaluation, we mix the target user’s authentication data with that of the passive attackers to assess the system’s ability to correctly accept the legitimate user while rejecting unauthorized attempts. As shown in Table~\ref{tab:ML-baseline}, all baselines achieve above 98.0\% BAC on our data. For comparison, EarID achieves a comparable BAC greater than 98.4\% across three key lengths, which is smaller only than that derived from MLP (99\%).

\subsection{Efficiency \label{sec:efficiency-evaluation}}
\noindent\textbf{Evaluation Platform.}
We evaluate the efficiency of EarID by emulating the authentication algorithm on two platforms: a mobile device and the Arduino Due, which serves as a proxy for the computational capacity of wireless earbuds. For the mobile device, we use a Google Pixel 5a smartphone equipped with a Snapdragon 765G processor, featuring 8 CPU cores running at 1.8 GHz and operating on Android 14. All programs, including EarID and baseline classifiers, are executed via Termux Android terminal emulator \cite{termux}. For the earbud-equivalent platform, we use an Arduino Due featuring an 80 MHz AT91SAM3X8E chip, due to the limited availability of open earable computing platforms.

\begin{figure}
    \centering
   \includegraphics[width=\linewidth]{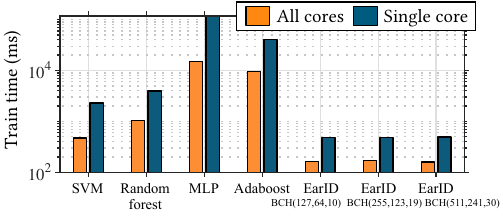}
    \caption{Baseline classifier training vs. EarID mobile enrollment time.}
    \label{fig:time}
\end{figure}

\smallskip
\noindent\textbf{Enrollment Time on Mobile.}
To emulate EarID’s enrollment process, one user’s enrollment data and population-level feature distribution are loaded onto the smartphone. When using all available cores, the computation time for EarID with key length 127/255/511 is approximately 155/160/178~ms. When restricted to a single core, this time increases to 491~ms.
With the same device and training strategy stated above, machine learning classifiers generally require longer training time than EarID, as shown in Fig.~\ref{fig:time}. With all cores, they range from 475~ms (SVM) to 14,824~ms (MLP), about 3x to 90x slower than EarID. The performance gap widens under single-core conditions, where classifiers need between 2,286~ms and 116,579~ms, i.e. up to 4x and 230x slower. This strong advantage in computational efficiency makes EarID suitable for personal on-device applications.

\smallskip
\noindent\textbf{Key Extraction Time on Earbuds.}
We then measure our key extraction time on the Arduino Due, representing the comparable (or even lower) computing capability of commercial earbuds. Since only feature extraction and binarization are required at the earbuds during authentication, our emulations report the required time for these steps. For BCH(127,64,10), 145~ms feature extraction time and 40~ms binarization result in a total 185~ms execution time. By extending key length to 255/511, it follows the same feature extraction but exhibit more bit extraction. Therefore, with the same feature extraction time (145~ms), BCH(255,123,19) and BCH(511,241,30) have longer key extraction time of 226~ms and 322~ms, respectively. Since the wireless module on earbuds have the coding and encryption capability for Bluetooth communication, no extra time is introduced to leverage this module for commitment.

\subsection{Robustness}
Robustness of our proposed EarID is evaluated through: resistance to unintended falsely triggered attempts (FT), cross-day and cross-session performances. The corresponding metrics include the falsely triggered acceptance rate, bit errors under cross-day and cross-session evaluations. We do not conduct extra experiments on in-body noise because the in-body noise can be picked up during the ECS process and our system is designed to be immune to it.

\smallskip
\noindent \textbf{Resistance to Unintended FT Attempts.} False triggering occurs when the earbuds are not intentionally worn by the legitimate user. Common conditions include being placed on a table, held in hand, or covered by cloth. Ideally, the authentication system should reject these attempts. To evaluate this, we deliberately collected the FT samples (40) and mixed them with the legitimate user attempts. The results, shown in Table~\ref{tab:earid-result}, indicate that all EarID configurations can effectively reject FT attempts.

\begin{figure}
    \centering
    \includegraphics[width=\linewidth]{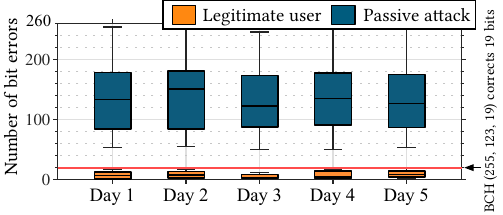}
    \caption{Cross-day evaluation: bit error of legitimate user vs. passive attacker vs. error correction from BCH(255,~131,~19).}
    \label{fig:cross-day}
\end{figure}

\smallskip
\noindent \textbf{Cross-Day Evaluation.} To evaluate long-term stability, we asked participants to wear the test earbuds, performing five authentication attempts per day for five successive days. Here we analyze the range of bit errors to evaluate system margin. As shown in Fig.~\ref{fig:cross-day}, using a moderate BCH(255,123,19), legitimate users typically exhibit 0–20 bit errors, with only 2 of 40 trials exceeding the correction threshold by 1 bit on Day 5. In contrast, passive attack samples show significantly higher errors, ranging from 50 to 255 bits. While the slight increase in error over time may suggest subtle aging effects, the large margin between legitimate users and attackers enables dynamic thresholding to maintain 100\% accuracy, e.g. with a 20-bit correction. Moreover, the lightweight design allows quick re-enrollment if the ECS key expires.

\begin{figure}
    \centering
    \includegraphics[width=\linewidth]{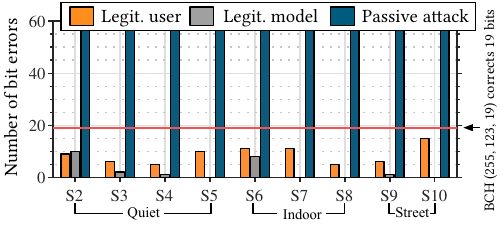}
    \caption{Cross-session evaluation: bit error of legitimate user/ear model vs. passive attacker vs. error correction from BCH(255,~131,~19).}
    \label{fig:cross-session}
\end{figure}

\smallskip
\noindent \textbf{Cross-Session Evaluation.}
For cross-session stability within short time frames, participants first enrolled themselves in the initial session (S1). Subsequently, participants repeatedly re-wore the earbuds for 9 times (S2-S9) and tested EarID each time while simulating noise conditions using a speaker and SPL meter in 20 minutes. Additionally, silicon ear model data were collected for reference in the same environment. An illustration of bit error performance in the cross-session evaluation for a human user and an ear model is shown in Fig~\ref{fig:cross-session}. For cross-session stability within short time frames, participants first enrolled themselves in the initial session (S1). They then repeatedly re-wore the earbuds in 9 follow-up sessions (S2-S9), and tested EarID under different noise conditions displayed by a smartphone: Quiet, Indoor, and Street, refer to Table~\ref{tab:audio_params}. For reference, silicon ear model data were also collected under the same conditions. As shown in Fig~\ref{fig:cross-session}, the cross-session evaluation demonstrates EarID's robust performance across multiple insertions. Both legitimate users and silicon ear models consistently show bit errors below the 19-bit correction threshold (red line) of BCH(255,123,19). In contrast, passive attacks exhibit significant higher bit errors across all sessions, making them easily to detect by EarID.

\begin{table*}[t]
\centering
\caption{Comparison of Related Earable/Head-Wearable Authentication Designs.}
\label{tab:related-work}
\scriptsize %
\begin{threeparttable}
    \resizebox{0.99\linewidth}{!}{%
        \begin{tabular}{llllcl}
        \toprule
        \textbf{Study} & \textbf{Purpose} & \textbf{Type} & \textbf{Auth Model} & \textbf{Subjects} & \textbf{Accuracy (Conditions)} \\ 
        \midrule
        EarGate \cite{EarGait} & Acoustic footstep & Phy & SVM & 31 & 97.3\% \\
        ToothSonic \cite{ToothSonic} & Tooth gesture & Beh & Neural Net & 25 & 95\% \\
        Voice-in-Ear \cite{Voice-in-ear} & In-ear voice & Phy & GMM-UBM + AdaBoost & 23 & 96.3\% \\
        Earprint \cite{EarPrint_OE} & TEOAE & Phy & LDA + Pearson Dist. & 54 & 99.4\% \\
        F2Key \cite{F2Key} & Acoustic mouth & CR & cGAN + Siamese NN & 26 & 95.3\%$\sim$99.9\% \\
        SonicID \cite{SonicID} & Ultrasound face & Phy & ResNet & 40 & 97.44\% (x-ses / days) \\ 
        \midrule
        Inaudible ECS \cite{2018inaudibleEar} & Hybrid ECS & Phy & LDA & 25 & 98.6\% (A-x-ses), 99.9\% (I-cont.) \\
        EarDynamic \cite{2021eardynamic} & Dynamic ECS & Phy & Classifier Boosting & 24 & 93\% \\
        HeadFi \cite{2021headfi} & ECS device& Phy & SVM & 27 & 95\% \\
        EarEcho \cite{2019earecho} & ECS classification & Phy & DT/SVM/MLP/etc & 20 & 95.1\% (x-ses), 97.6\% (cont.) \\
        Bilateral ECS \cite{2022bilateral} & Bilateral ECS & Phy & SVM & 52 & 99.6\% \\ 
        \midrule
        \cellcolor{darkgray!10}\text{EarID (ours)} & \cellcolor{darkgray!10}\text{ECS over wireless earbuds} & \cellcolor{darkgray!10}\text{Phy} & \cellcolor{darkgray!10}\text{Key extraction} & \cellcolor{darkgray!10}\text{44} & \cellcolor{darkgray!10}\text{98.7\% (A-cont / x-ses / x-day)}\\ 
        \bottomrule
        \end{tabular}
    }
    \begin{tablenotes}
        % \scriptsize % 
        \item Phy: physiological; Beh: behavioral; CR: challenge response; 
        \item TEOAE: Transient-Evoked Otoacoustic Emissions;
        \item x-ses/days: cross session and days; cont.: continuous authentication.
        \item A: audible band; I: inaudible band.
    \end{tablenotes}
\end{threeparttable}
\vspace{-5mm}
\end{table*}

\subsection{Security}
We assess EarID's security by evaluating its resistance to various attackers outlined in the threat model. The metric used for this evaluation is the attack success rate (ASR) for each attack, which indicates the proportion of successful bypass attempts by a specific type of attacker, granting them access as the legitimate user. 

\smallskip
\noindent \underline{\circled{1}: Passive Attack.} Passive attack samples are selected from other user's data other than the legitimate users. Specifically, in each independent test, we randomly designate 30 participants in the dataset as the gallery dataset, providing population-level statistics and feature probability distribution. Meanwhile. the remaining 14 users enroll with their own ECS data. These enrolled users attempt (a) legitimate authentication on their own account and (b) passive attacks by attempting to access another user’s account. The result is aggregated in 20 independent trails. Table~\ref{tab:earid-result} shows that the EarID across all three BCH settings successfully defends against such intrusions, with passive attacker ASR (P-ASR) consistently below 1\%.

\smallskip
\noindent \underline{\circled{2}: Synthetic ECS Attack.} To simulate synthetic ECS attacks, we wear the wireless earbuds on a silicon ear model (Fig.~\ref{fig:UI}(a)) to collect 40 synthetic samples (S-attacks). In each independent test, we randomly re-group the user similar to passive attack, and use S-attack samples to to attempt authentication into enrolled accounts. The resulting S-ASR remains low and comparable to P-ASR, indicating that the system detects the ear model as an unauthorized source.

\smallskip
\noindent \underline{\circled{3}: Universal ECS Attack.} Universal ECS attacks target users with ECS features that are common or non-distinct, when attacker can exploit a public gallery dataset to construct generic ECS patterns. The attacker selects the most common ECS feature from the public dataset and uses it to attempt access. While this approach mimics human data more closely than synthetic ear models, it still fails against EarID. This is because EarID explicitly selects high-biometric-information features (Eq.~\ref{eq:bioinfo}) during key extraction, while discarding common features. As a result, features used in universal attacks cannot match the legitimate user's feature mask, and U-ASR remains near 0\% shown in Fig.~\ref{fig:bit-error-eer}. 

\smallskip
\noindent \underline{\circled{4}: Key Guessing Attack.} Instead of random guess the extracted binary key, attackers may user their own EarID for key guessing, expecting higher success rate if EarID cannot emphasize each user's uniqueness and generate low-entropy keys. However, this strategy proves ineffective, which leads to a 0\% K-ASR and a FAR curve that hugs the x-axis in Fig.~\ref{fig:bit-error-eer}, confirming that such key guessing attempts fail completely.
\\

\section{User Hearing Experience} 
To incorporate human auditory perception into our evaluation, we employed Zwicker’s psychoacoustic annoyance (PA) model \cite{zwicker_PA} to quantify the subjective listening experience of the transmitted ECS signal. Using the SQAT toolbox \cite{SQAT}\footnote{The minimum audio duration required by the toolbox is 2 seconds. This criterion is met as both the 1s-MLS and 1s-chirp signals are played twice (totaling 2 seconds) for data aggregation.}, we obtained PA values of 14.8 for the chirp signal and 10.7 for the MLS signal (parameters detailed in Table~\ref{tab:audio_params}). For context, 20 commonly used “notification” and “alarm” sound effects sourced online \cite{mixkit} exhibited an average PA value of 15.4.
Additionally, a Mean Opinion Score (MOS) assessment was conducted to capture participants' subjective experiences of the ECS sounds. Participants rated the sounds on a 5-point scale (1: Very Annoying, 5: Very Pleasant). The average MOS scores for the chirp and MLS signals were 3.2 and 2.2, respectively. Although the MOS results contrast with the psychoacoustic annoyance (PA) values, they indicate that the ECS signals effectively notify the user while maintaining an auditory experience within an acceptable range. It is worth noting that ECS can operate in the inaudible frequency band \cite{2018inaudibleEar}. However, this approach has been reported to be suitable primarily for continuous authentication, which falls outside the scope of this study, as our focus is on improving efficiency.

\section{Discussion \label{sec:discussion}}
\noindent{\textbf{Impact of Bandpass Filtering.}}
As discussed in Section~\ref{sec:client-design}, the bandpass filter plays a crucial role in mitigating additive and convolutional noise. We evaluated the impact of different passband parameters on EarID-BCH(255,123,19) by testing (0 Hz - 8 kHz), (2 kHz - 4 kHz), (2 kHz - 8 kHz), (2 kHz - 16 kHz), and (8 kHz - 16 kHz). The corresponding BAC results were $68.5\%$, $97.3\%$, $98.7\%$, $97.9\%$, and $96.3\%$. These results indicate that low frequencies below 2 kHz adversely affect system performance. Also, large bandwidth does not always enhance accuracy in EarID system. 

\smallskip
\noindent{\textbf{Impact of Biometric-Informative Masking.}}
The purpose of biometric-informative masking is to filter out cepstral features that lack uniqueness or exhibit high intra-subject variability. To study its impact, we excluded this module and generate ECS code for matching. Without biometric-informative masking, the FAR dropped to $0\%$, but the FRR soared to $86.9\%$. This demonstrates that the proposed feature selection module effectively identifies useful cepstral features, significantly improving EarID system performance.

\smallskip
\noindent{\textbf{Selection of Error Correction Code.}}
{ We tested only BCH as the error correction code (ECC) for EarID, while other options including Reed-Solomon (RS) \cite{reed-solomon} and Low-Density Parity-Check (LDPC) \cite{ldpc} codes are more complex to perform. This selection can be optimized in future work, as EarID relies on ECC to tolerate random bit errors caused by variations in the user's ear canal and the way earbuds are worn. It is noteworthy that the \textit{flexible} (yet sufficient) error-correction capacity, rather than a high error-correction capacity, benefits EarID. Additionally, complexity is an important consideration when implementing ECC on earable devices. Nevertheless, ECC is fundamental for wireless systems and might be utilized without incurring extra costs.}

\smallskip
\noindent{\textbf{System Update.}}
During the warm-up stage of ECS-based biometric authentication, it is essential to collect a sufficiently annotated gallery dataset of biometric features to develop an accurate and robust system. Continuous data collection from an increasing number of consenting users facilitates system updates. When an update is necessary, all users opt to re-enroll based on the new gallery dataset to enhance security. In this context, EarID's learning-free and low-complexity design is advantageous for system updates and can cease once the population distribution converges.

\smallskip
\noindent{\textbf{Potential Extensions.}}
The proposed EarID system can be considered a promising biometric authentication solution for personal mobile device paired with earables. However, the fundamental challenge of the new biometric modality studies is still the lack of a public ECS dataset. To address this issue, we will release details of our hardware and software platforms to support ECS data collection for new research. Additionally, our proposed EarID can also be improved in the future. For example, using index-based IoM-hashing \cite{iomhashing} can potentially be more efficient compared with the current random projection-based BioHashing. Lastly, the new ECS representations proposed in this work can effectively capture in-ear sensing information, potentially extending their application to other human in-ear sensing research, for example, detecting the ear canal deformation similar to EarDynamic \cite{2021eardynamic} with EarID's binary representation.

\section{Related Work \label{sec:related-work}}
\noindent{\textbf{Wearable Authentication Methods.}} Listed in Table.~\ref{tab:related-work}, several related works achieve high authentication accuracy using non-ECS (Ear Canal Scanning) acoustic sensing modalities. For example, EarGate \cite{EarGait}, ToothSonic \cite{ToothSonic}, and Voice-in-Ear \cite{Voice-in-ear} authenticate users by passively recording in-body sounds, while Earprint \cite{ear-occlusion-effect}, F2Key \cite{F2Key}, and SonicID \cite{SonicID} actively emit acoustic signals to drive authentication. Notably, SonicID stands out for its evaluation across sessions and days, demonstrating long-term consistency.

\smallskip
\noindent{\textbf{ECS-Based Authentication Methods.}} Among ECS-based authentication designs, different approaches emphasize specific aspects, such as inaudible ECS \cite{2018inaudibleEar}, dynamic deformation \cite{2021eardynamic}, device modification \cite{2021headfi}, classifier ensemble \cite{2019earecho}, and bilateral ear \cite{2022bilateral}. These works commonly use classifiers to perform the authentication task and are reproduced using our data, as shown in Table.~\ref{tab:related-work}.

\smallskip
\noindent{\textbf{Our Contribution.}} Comparing with existing authentication method, our unique contributions include:
\begin{itemize}
    \item Practical biometric authentication by fully extracting key on earbuds and verifying at mobile.
    \item No raw ECS data transmission to mobile device to mitigate biometric stealing over insecure channels.
    \item Efficiency in enrollment and ability to authenticate on resource-limited wireless earbuds.
\end{itemize}

\section{Conclusion \label{sec:conclusion}}
In conclusion, we present EarID, a learning-free biometric authentication system that leverages ear canal scanning (ECS) key extraction. Our target is to implement the ECS authentication directly on wireless earbuds, in order to avoid raw biometric data transmission over insecure wireless environment during authentication. Unlike existing methods that rely on ML classifiers, we design the lightweight learning-free key extraction approach that can execute on wireless earbuds, condensing the user-specific unique features into binary keys. Furthermore, the derived key is integrated into the fuzzy commitment scheme for secure earbuds-to-mobile authentication. Our extensive evaluations demonstrate that EarID achieves high authentication accuracy up to 98.7\%, which is comparable with classifiers but is faster in mobile enrollment (160~ms) and enables earbuds key extraction (226~ms). EarID also shows strong resilience against false triggering and multiple threat models, with all false acceptance rate lower than 1\%. These results suggest that EarID is a practical and secure solution for seamless biometric authentication on wireless earbuds.
\newpage

\bibliographystyle{unsrt}
\bibliography{ref.bib}

\end{document}